\documentstyle[12pt]{article}
\newcommand{\be}{\begin{equation}}
\newcommand{\ee}{\end{equation}}
\newcommand{\bea}{\begin{eqnarray}}
\newcommand{\eea}{\end{eqnarray}}
\newcommand{\ba}{\begin{array}}
\newcommand{\ea}{\end{array}}
\newcommand{\beas}{\begin{eqnarray*}}
\newcommand{\eeas}{\end{eqnarray*}}
\newcommand{\bes}{\begin{equation*}}
\newcommand{\ees}{\end{equation*}}

\begin{document}
\title{\bf Holograghic Brownian motion in $2+1$ dimensional hairy black holes}
\author{J. Sadeghi$^{a}$ \thanks{Email: pouriya@ipm.ir}\hspace{1mm} B. Pourhassan$^{b}$\thanks{Email: b.pourhassan@umz.ac.ir}, and F. Pourasadollah$^{a}$ \thanks
{Email: f.pourasadollah@gmail.com}\\
$^{a}${\small {\em  Sciences Faculty, Department of Physics, Mazandaran University,}}\\
{\small {\em P.O.Box 47416-95447, Babolsar, Iran}}\\
$^{b}${\small {\em Department of Physics, Damghan
University, Damghan, Iran}}}
 \maketitle
\begin{abstract}
In this paper, we investigate the dynamics of a heavy quark under the plasmas correspond to three dimensional hairy black holes. We utilize the AdS/CFT correspondence to study the holographic Brownian motion of this particle in different kinds of hairy black holes. For uncharged black hole at low frequency limit we derive analytic expressions for correlation functions and response functions and verify that the fluctuation-dissipation theorem holds in the presence of a scalar field in the metric background. In the case of charged black hole, we think that the results are similar to what derived for uncharged black hole.
\end{abstract}
\section{Introduction}
Heavy ion collisions experiments at RHIC are believed to create strongly coupled quark gluon plasma (sQGP) \cite{P1}-\cite{P3}. The QGP is a phase of QCD that thought to be very similar to the plasma of N = 4 super Yang Mills theory at finite temperature. One of the current challenges in theoretical particle physics is to compute properties of this strongly coupled plasma. The AdS/CFT correspondence \cite{P4}-\cite{P7} has led to many profound insights in to the nature of strongly coupled gauge theories. This gauge/gravity duality provide the possibility of computing some properties of QGP \cite{P8}, \cite{Pour1}, \cite{Pour2} and \cite{P9}. QGPs contains quarks and gluons, like hadrons, but dislike hadrons,  the mesons and baryons lose their identities and dissolve into a fluid of quarks and gluons. A heavy quark immersed in this fluid, undergoes the Brownian motion  \cite{P10}-\cite{P13} at finite temperature. The AdS/CFT correspondence can be utilized to investigate the Brownian motion of this particle. In the context of this duality, the dual statement of the quark in QGP corresponds to the end point of an open string that extends from the boundary to the black hole horizon. The black hole environment excites the modes of string by the Hawking radiation. It was found that, once these modes are quantized, the end point of string at the boundary shows the Brownian motion which is described by the Langevin equation \cite{P11}-\cite{P13}.\\ In the formulation of AdS/CFT correspondence, fields of gravitational theory would be related to the corresponding boundary theory operators \cite{P6} and \cite{P7}, such as their boundary value should couple to the operators. In this way, instead of using the boundary field theory to obtain the correlation function of quantum operators, one can determine these correlators by the thermal physics of black holes and use them to compute the correlation functions. For different theories of gravity one can make an association with various plasmas in the boundary. In this paper we follow different works \cite{P15}-\cite{P20} to investigate the Brownian motion of a particle in two dimensional plasma which its gravity dual is described by three dimensional hairy metric background \cite{P21}-\cite{P28}. We obtain the solutions to the equation of motion of uncharged hairy black holes by using matching teqnique at low frequency limit. We utilize these solutions in investigating the Brownian motion of the particle. We receive some expersion for responce functions and correlation functions and show that the fluctuation-dissipation theorem holds in the presence of a scalar field in the metric background. For the case of charged black hole we make some comments.\\
This paper is structured as follows: in section 2, we give some review for different kind of three dimensional
hairy black holes. Section 3 is assigned to look for a holographic realization of Brownian motion in the boundary and bulk side of theory. We study the Holographic Brownian motion in hairy black holes and the fluctuation-dissipation theorem in the presence of a scalar field in the metric back ground in section 4. In section 5, we summarize our works in this paper and make some comments about our results and close with conclusions.
\section{Hairy black holes in $2+1$ dimension}
There is already a huge amount of literature on the subject of gravity coupled with a scalar field \cite{P21}-\cite{P28}. Black hole solutions in such theories are known as hairy black holes. In this paper, we are interested to study the black hole solution in an Einstein-Maxwell-scalar gravity with a nonminimally coupled scalar field in (2 + 1) dimensions \cite{P21}-\cite{P23}. The action reads,
\begin{equation}
I=\frac{1}{2}\int d^{3}x
\sqrt{-g}[R+g^{\mu\nu}\nabla_{\mu}\phi\nabla_{\nu}\phi-\xi R \phi^{2}-2V(\phi)-\frac{1}{4}F_{\mu\nu}
F^{\mu\nu}],
\end{equation}
where $\xi$ is a coupling constant between gravity and the scalar field which will be fixed be as
$\xi=\frac{1}{8}$ \cite{P22}.\\
A static, circularly solution to the above action which represents charged hairy black hole, can be written by,
\begin{equation}\label{eq 4}
 ds^{2}=-f(r)dt^{2}+ \frac{1}{f(r)} dr^{2}+ r^{2} d\psi^{2},
\end{equation}
where,
\begin{eqnarray}
f(r)=\frac{r^{2}}{l^{2}}-M+\frac{(\frac{Q^{2}}{6}-2M)B}{3r}-\frac{Q^{2}}{2}(1+\frac{2B}{3r})\ln(r),
\end{eqnarray}
$Q$ is the electric charge, $l$ is the integration constant and $B$ is related to the scalar field as,
\begin{equation}
\phi(r)=\pm\sqrt{\frac{8B}{r+B}}.
\end{equation}
When $B=0$, the scalar field $\phi$
vanishes, and the system becomes the Einstein-Maxwell-AdS theory. The solution is known as static charged BTZ black hole \cite{P28}.
\begin{equation}
 f(r)=\frac{r^{2}}{l^{2}}-M-\frac{Q^2}{2}\ln(r),
\end{equation}
where $M$ is the mass of BTZ black hole. The horizon radius in the terms of black hole mass and charge obtained as,
\begin{equation}
  r_{+}=exp\left(-\frac{1}{2}Lambert W \left[\frac{-4exp(\frac{-4M}{Q^{2}})}{l^{2}Q^{2}}\right]-\frac{2M}{Q^{2}}\right).
\end{equation}
In the case of uncharged black hole we set $Q=0$ in equation (3), then this equation reduced to the following,
\begin{equation}
f(r)=\frac{r^{2}}{l^{2}}-M-\frac{2BM}{3r}=\frac{(r-r_{+})(r-r_{1})(r-r_{2})}{3rl^{2}},
\end{equation}
where,
\begin{equation}
r_{+}=\frac{1}{3}Y(M,B)+\frac{Ml^{2}}{Y(M,B)},
\end{equation}
is the black hole horizon with,
\begin{equation}
 Y(M,B)=[9BMl^{2}+3\sqrt{-3M^{3}l^{6}+9B^{2}M^{2}l^{4}}]^{\frac{1}{3}}.
\end{equation}
In the relation (7) $r_{1}$ and $r_{2}$ are given by,
\begin{eqnarray}
r_{1}=-\frac{1}{2}[\frac{1}{3}Y(M,B)+\frac{Ml^{2}}{Y(M,B)}]+i\frac{\sqrt{3}}{2}
[\frac{1}{3}Y(M,B)-\frac{Ml^{2}}{Y(M,B)}]=r_{2}^{*}.
\end{eqnarray}
We can derive the mass and temperature of black hole in the terms of entropy $s$ as following,
\begin{equation}
M=\frac{3s^{3}}{16 \pi^{2}l^{2}(8\pi B+3s)}\qquad ,\qquad T=\frac{9s^{2}(4\pi B+s)}{8\pi^{2}l^{2}(8\pi B+3s)^{2}},
\end{equation}
where $s=4\pi r_{+}$. In the special case of $M=\frac{3B^{2}}{l^{2}}$, where the self interacting scalar vanishes, we receive to the conformal black hole with \cite{P24},
\begin{equation}
f(r)=\frac{(r-2B)(r+B)^{2}}{rl^{2}},
\end{equation}
which yields to the following black hole mass and temperature,
\begin{equation}
M=\frac{3s^{2}}{64 \pi^{2}l^{2}}\qquad ,\qquad T=\frac{3s}{32\pi^{2}l^{2}}.
\end{equation}
\section{Holographic Brownian motion}
\subsection{Dictionary of Brownian motion in the boundary}
 In the field theory or boundary side of AdS/CFT story for Brownian motion, a mathematical description of this motion
 is given by the Langevin equation  \cite{P11}-\cite{P13} which has the generalized form as,
  \begin{equation}
\dot{p}(t)=-\int_{-\infty}^{t}\gamma(t-\acute{t})p(\acute{t})+R(t)+K(t)\,,
\end{equation}
where $p$ is momentum of Browniam particle. The terms in the right-hand side of (14) correspond
to the friction, random and external force respectively and $\gamma(t)$ is the kernel function. At a first
approximation, one can assume the following averages for random force:
 \begin{equation}
\left<R(t)\right>=0,\qquad\qquad \left<R(t)R(\acute{t})\right>=\kappa_{0}\delta(t-\acute{t}),
\end{equation}
where $\kappa_{0}$ is a constant which, due to the fluctuation-dissipation, is related to the friction coefficient $\gamma_{0}$ through,
\begin{equation}
  \kappa_{0}=\frac{\gamma_{0}}{2mT}.
\end{equation}
This comes from the fact that the frictional and random forces have the same origin at the microscopic level (collision with the fluid constituents).\\
 The time evolution of displacement square for long times (or low frequencies) can be derived by computing
the two-point correlation function, so it results to the following relation,
 \begin{equation}
  \langle s(t)^{2}\rangle=\langle[x(t)-x(0)]^{2}\rangle\approx\ 2Dt,\qquad\qquad \mathrm{for} \qquad
  t\gg\frac{1}{\gamma_{0}},
 \end{equation}
 where $D=\frac{T}{\gamma_{0}m}$ is the diffusion constant and we have assumed that $\langle m\dot{x}^{2}\rangle=T$.
\\After applying the Fourier transformation for relation (14) one can obtain,
 \begin{equation}
 p(\omega)=\frac{R(\omega)+K(\omega)}{\gamma[\omega]-i\omega}.
\end{equation}
If we take the statical average of (18), then we have,
\begin{equation}
\left<p(\omega)\right>=\mu(\omega)K(\omega),\qquad\qquad \mu(\omega)\equiv\frac{1}{\gamma[\omega]-i\omega},
\end{equation}
$\mu(\omega)$ is known as the admittance. So we can determine this quantity by measuring the response $p(\omega)$ to an external force.\\
The power spectrum $I_{O}(\omega)$, is defined for a quantity $O(t)$ as,
\begin{equation}
I_{O}(\omega)=\int_{-\infty}^{\infty}dt \langle O(t_{0})O(t_{0}+t)\rangle e^{i\omega t},
\end{equation}
 and it is related to the two-point function because of the Wiener-Khintchine theorem,
 \begin{equation}
\langle O(\omega)O(\acute{\omega})\rangle=2\pi \delta(\omega+\acute{\omega})I_{O}(\omega).
\end{equation}
For the case without external force ,$i.e.,K=0$, by using relation (18) and after some algebra, one gets that,
\begin{equation}
\kappa(\omega)=I_{R}(\omega)=\frac{I_{p}(\omega)}{|\mu(\omega)|^{2}}.
\end{equation}
This will be important for checking the validity of the fluctuation-dissipation theorem.
\subsection{Brownian motion in the bulk}
An external quark on the boundary theory can be realized as the endpoint of an open string which is hanged from the
boundary and dips into the black hole horizon. The dynamics of this string is governed by the Nambu-Goto action \cite{P15}-\cite{P18},
\begin{equation}\
 S_{NG}=-\frac{1}{2\pi\alpha^{\prime}}\int d\tau d\sigma
 \sqrt{-\det g_{_{ab}}}\,,
\end{equation}
where $g_{ab}= G_{\mu \nu}\partial_{a}X^{\mu}\partial_{b}X^{\nu}$ denotes the induced metric on the worldsheet. We set $\tau=t$ and $\sigma=r$ to work in the static gauge. Our string embedding is then given by $X^{\mu}(t,r)=(t,r,\psi(t,r))$. This string stretches along the r direction and has small fluctuations in the transverse direction $\psi(t,r)$. For $\psi(t,r)=c$, one can check that this yields to a trivial solution which correspond to a quark in equilibrium in the thermal bath. In this case the mass of particle can be easily computed from the tension of the string,
\begin{equation}
 m =\frac{1}{2\pi \alpha^{\prime}}\int_{r_{+}}^{r_{b}} dr \sqrt{-g_{tt}g_{rr}}=\frac{1}{2\pi
 \alpha^{\prime}}(r_{b}-r_{+})\approx\frac{1}{2\pi \acute{\alpha}}r_{b},\qquad for \quad   r_{b}\gg r_{+},
\end{equation}
where $r_{b}$ is the position of boundary and $r_{+}$ is defiend as horizon radius of black hole.\\
 If the scalar $\psi(t,r)$ do not fluctuate too far from its equilibrium values ($\psi(t,r)=0$),
 we can expand the Nambu-Goto action up to quadratic order in perturbations \cite{P3} as,
\begin{equation}\
 S_{NG}\approx-\frac{1}{4\pi\alpha^{\prime}}\int dt\,dr\:r^{2}\left[f(r)\psi'^{2}-\frac{1}{f(r)}\dot{\psi}^{2}\right],
\end{equation}
where $\psi'\equiv \partial_{r}\psi$ and $\dot{\psi}\equiv \partial_{t}\psi$.\\ If we consider an external force on the Brownian particle as in (14), we can obtain the admittance from the response of particle to this force. This situation can be easily yielded by turning on world-volume electric field on the flavor D-brane at $r=r_{b}$. Since the endpoint of sting is charged, this is equal to add a boundary term to the Nambu-Goto action $S=S_{NG}+S_{EM}$, where
\begin{equation}\
 S_{EM}=\oint(A_{t}+A_{\psi}\dot{\psi})dt.
\end{equation}
This will affect on the motion of the external quark on the boundary and will not play any role for the string dynamics on the bulk. We leave this part of action for now but we will need it later.\\
 The Nambu-Goto action (25) near the horizon limit $r\sim r_{+}$ becomes,
\begin{equation}\
 S_{NG}\approx-\frac{1}{4\pi\alpha^{\prime}}\int dt\,dr_{*}\:r_{+}^{2}\left[\psi'^{2}-\dot{\psi}^{2}\right].
\end{equation}
Here, the primes stand for derivatives with respect to the tortoise coordinates $r_{*}$, which is defined by,
\begin{equation}
dr_{*}=\frac{dr}{f(r)}
\end{equation}
Thus, the equation of motion is then,
\begin{equation}
(\partial_{ r_{*}}^{2}-\partial_{t}^{2})\psi=0,
\end{equation}
which show that in the near horizon limit $\psi$ behaves likes massless Klein-Gordon scalars. If we define
$\psi(t,r)=e^{-i\omega t}g_{\omega}(r) $, then the two independent solutions to the equation (29) are,
\begin{eqnarray}
 \psi^{out}(t,r)&=& e^{-i\omega t}g^{out}(r)\sim e^{-i\omega (t-r_{*})} \\
 \psi^{in}(t,r)&=& e^{-i\omega t}g^{in}(r)\sim e^{-i\omega (t+r_{*})}.
 \end{eqnarray}
Following the standard quantization of quantum fields in curved spacetime, we receive to a mode expansion  of the form,
 \begin{equation}
 \psi(t,r)=\int_{0}^{\infty}\frac{d \omega}{2\pi}[a_{\omega}u_{\omega}(t,r)+a_{\omega}^{\dag}u_{\omega}(t,r)^{*}],
 \end{equation}
 with
 \begin{equation}
 u_{\omega}(t,r)=\eta\left[g^{out}(r)+\delta\:g^{in}(r)\right]e^{-i\omega
 t}\qquad,\qquad\left[a_{\omega},a_{\acute{\omega}}^{\dag}\right]=2\pi\delta(\omega-\acute{\omega})\,.
 \end{equation}
 $\eta$ and $\delta$ are constants that are found by requiring the normalization of modes through the standard
 Klein-Gordon inner product and the boundary condition  at $r=r_{b}$ respectively. The string modes satisfy the Bose-Einstein distribution:
 \begin{equation}
<a_{\omega}^{\dag}a_{\acute{\omega}}>=\frac{2\pi\delta(\omega-\acute{\omega})}
{e^{\beta\omega}-1}\,,
\end{equation}
 with $\beta=\frac{1}{T}$. Using this and the mode expansion given in (33), we compute,
\begin{eqnarray}
 \left<:\Psi(t)\Psi(0):\right>=\left<:\psi(t,r_{b})\psi(0,r_{b}):\right> =\int_{0}^{\infty}\frac{d
 \omega}{2\pi}\frac{1}{e^{\beta\omega}-1}\left[u_{\omega}(t,r_{b})u_{\omega}(0,r_{b})^{*}+
 u_{\omega}(t,r_{b})^{*}u_{\omega}(0,r_{b})\right],\nonumber\\
                      = \int_{0}^{\infty}\frac{d \omega}{2\pi} \frac{2 |\eta|^{2}\cos(\omega
                      t)}{e^{\beta\omega}-1}|g^{out}(r_{b})+\delta\,g^{in}(r_{b})|^{2}.\qquad\qquad\quad\;\;\,
 \end{eqnarray}
 From the above relation we can easily derive the general form of the momentum correlator,
  \begin{eqnarray}
\left<:p(t)p(0):\right>=-m^{2}\partial_{t}^{2}\left<:\psi(t,r_{b})\psi(0,r_{b}):
\right>,\qquad\qquad\qquad\qquad\;\,\,\nonumber\\
                      = \int_{0}^{\infty}\frac{d \omega}{2\pi} \frac{2m^{2}\omega^{2} |\eta|^{2}\cos(\omega
                      t)}{e^{\beta\omega}-1}|g^{out}(r_{b})+\delta\,g^{in}(r_{b})|^{2}.
 \end{eqnarray}
 Eventually, the displacement square can be obtained from the relation (35) as following,
  \begin{eqnarray}
S^{2}(t)=\left<:[\Psi(t)-\Psi(0)]^{2}:\right>= \int_{0}^{\infty}\frac{d \omega}{2\pi} \frac{8
 |\eta|^{2}\sin^{2}(\frac{\omega t}{2} t)}{e^{\beta\omega}-1}|g^{out}(r_{b})+\delta\,g^{in}(r_{b})|^{2}.
  \end{eqnarray}
 \section{Brownian motion in Hairy black holes}
 \subsection{String dynamics in Hairy black holes and the response function}
By having the knowledge of last sections in hand, we are now able to realize the Brownian motion of a particle
moving on the boundary of hairy black holes. The dual state of such a particle is an open string which extends from
 the boundary to the horizon of the hairy black holes.  We can obtain the dynamics of this string through the
 Nambu-Goto action (25). The equation of motion derived from this action is,
\begin{equation}
\frac{\partial}{\partial \rho}\left[\rho^{2}f(\rho)\frac{\partial g_{\omega}(\rho)}{\partial \rho}\right]+
\frac{4B^{2}\rho^{2}\omega^{2}}{f(\rho)}g_{\omega}(\rho)=0\,,
\end{equation}
where we have defined $\rho=\frac{r}{r_{h}}$ and used $\psi(t,r)=e^{-i\omega t}g_{\omega}(r)$. In general, one can
check that it is not possible to solve the above equation analytically for any kind of $f(r)$ defined in the section 2. However, we can employ a low frequency approximation by means of the matching technique\cite{P15}, \cite{P18}-\cite{P20}. To find the
solutions in this way, we consider three regimes:\\
(I) the near horizon solution $(\rho\sim1)$ for arbitrary $\upsilon$ (where $\upsilon=\frac{l^{2}\omega}{r_{h}}$),\\
(II)the solution for arbitrary $\rho$ in the limit $\upsilon\ll 1$ and\\
(III)the asymptotic $\rho\rightarrow\infty$ solution for arbitrary $\upsilon$,\\
and find the approximate solutions for each of regimes then we match them to leading order in $\upsilon$. We implement the above method for each kind of uncharged and charged hairy black holes separately.
In this section, we begin our discussion with the special case of conformal black hole, then we continue on uncharged black hole with the general mass. We study the charged hairy black hole at the end of this section.
\subsubsection{Uncharged black hole with the special mass (Conformal black hole)}
 In the conformal black hole, with $f(r)$ defined through relation (12), we expect two solutions in regime (I) of the form of relations (31),(32). In the tortoise coordinate system which utilized in those solutions, the $(t,r_{*})$ part of metric is conformally flat and the equation of motion near the horizon have a behavior similar to the wave equation in flat space. So, we choose solutions (31),(32), where the parameter $r_{*}$ has the following definition,
\begin{equation}
  r_{*}\sim \frac{2l^{2}}{9B}\ln(2B-r)
\end{equation}
near the horizon, $r=2B$, for conformal black hole. In fact in the near horizon regime, our equation of motion for this kind of black hole reduce to,
\begin{equation}
g_{\upsilon}(\rho)''+\frac{1}{\rho-1}g_{\upsilon}(\rho)'+ \frac{16\upsilon^{2}}{81(\rho-1)^{2}}g_{\upsilon}(\rho)=0\,,
\end{equation}
consequently independent solutions are obtained as,
\begin{equation}
g_{\upsilon}^{out/in}(\rho)=e^{\pm\frac{4i\upsilon}{9}\ln(\rho-1)}=1\pm\frac{4i\upsilon}{9}\ln(\rho-1)
+O(\upsilon^{2})\;.
\end{equation}
We can expand $g_{\upsilon}(\rho)$ as a power series in the $\upsilon$ for regime (II), then we have,
\begin{equation}
  g_{\upsilon}(\rho)=g_{\upsilon}^{(0)}(\rho)+\upsilon
  g_{\upsilon}^{(1)}(\rho)+\upsilon^{2}g_{\upsilon}^{(2)}(\rho)+...
\end{equation}
The first term can be derived from solving the following equation,
\begin{equation}
\frac{\partial}{\partial \rho}\left[\rho^{2}f(\rho)\frac{\partial g_{\upsilon}^{(0)}(\rho)}{\partial \rho}\right]=0\,.
\end{equation}
The general solution in this regime for any kind of $f(\rho)$ is given by,
\begin{equation}
  g_{\upsilon}^{(0)}=B_{1}+B_{2}\int\frac{d\rho}{\rho^{2}f(\rho)}.
\end{equation}
In the special case of conformal black hole we attain,
\begin{equation}
  g_{\upsilon}^{(0)}(\rho)=B_{1}+B_{2}\left[\frac{8}{9}\ln(2\rho+1)-\ln\rho+\frac{1}{9}\ln(\rho-1)-
  \frac{2}{3(2\rho+1)}\right].
\end{equation}
In the regime (III), one has $f(\rho)\rightarrow \frac{4B^{2}\rho^{2}}{l^{2}}$ for $\rho\rightarrow\infty$, then the equation (38) becomes,
\begin{equation}
\frac{\partial}{\partial \rho}\left[\rho^{4}\frac{\partial g_{\upsilon}(\rho)}{\partial \rho}\right]+\upsilon^{2}
g_{\upsilon}(\rho)=0\,.
\end{equation}
The general solution to the above equation can be found as a series expansion in $\frac{1}{\rho}$. For leading order terms we have,
\begin{equation}
g_{\upsilon}(\rho)=C_{1}(1+\frac{\upsilon^{2}}{2\rho^{2}})+C_{2}\frac{i\upsilon^{3}}{3\rho^{3}}+
O(\frac{1}{\rho^{4}})\;.
\end{equation}
In order to have the asymptotic solutions at low frequencies, we require to obtain the coefficients. So we expand (45) near the horizon and match the solution with (41) to find $B_{1}$ and $B_{2}$. From (41) and (45) it follows that,
\begin{equation}
B_{1}^{(out/in)}=1\mp\frac{4i\upsilon}{9}(8\ln3-2)\qquad\qquad,\qquad\qquad B_{2}^{(out/in)}=\pm4i\upsilon\;.
\end{equation}
Finally, expanding the solution in (45) for $\rho\rightarrow\infty$ yields,
\begin{equation}
  g_{\upsilon}^{(0)}(\rho)=B_{1}+B_{2}\left[\frac{8}{9}\ln2+\frac{-1}{12\rho^{3}}\right].
\end{equation}
By comparing equation (49) with (47) and using (48) we obtain,
\begin{equation}
  C_{1}^{(out/in)}=1\mp\frac{4i\upsilon}{9}(8\ln\frac{3}{2}-2)\qquad\qquad,\qquad\qquad
  C_{2}^{(out/in)}=\mp\frac{1}{\upsilon^{2}}.
\end{equation}
Thus, the asymptotic solutions, in the low frequency limit, and for corresponding modes of outgoing and incoming wavefunctions at the horizon, one can obtain
\begin{equation}
g_{\upsilon}^{(out/in)}(\rho)=1\mp\frac{4i\upsilon}{9}(8\ln\frac{3}{2}-2)(1+\frac{\upsilon^{2}}{2\rho^{2}})
\mp\frac{i\upsilon}{3\rho^{3}}+O(\frac{1}{\rho^{4}}).
\end{equation}
We now turn our attention to the case that an external force imposes on the Brownian particle. We can compute the admittance
from the response of particle to this force. As mentioned in section 2, in this case, because
 of the new boundary condition, we must modify our equation of motion as
 \begin{equation}
   \frac{\partial\mathcal{L}}{\partial \psi_{i}'}=F_{i} \quad,\qquad where\quad F_{i}=-(F_{it}+F_{ij}\dot{\psi}^{j}),
  \end{equation}
with $F_{it}= \partial_{i}A_{t}-\partial_{t}A_{i}$ and $ F_{ij}= \partial_{i}A_{j}-\partial_{j}A_{i}$. For a general
metric background defined in the relation (2) we get,
\begin{equation}
  F_{t\psi}=\frac{r_{h}f(\rho)\rho^{2}\partial_{\rho}\psi}{2\pi \acute{\alpha}}
  \qquad\qquad\mathrm{at}\qquad\rho=\rho_{b}.
\end{equation}
 The general solution for $\psi$ is the combination of ingoing and outgoing modes. In the semiclassical approximation, because of Hawking radiation \cite{P29},\cite{P30}
 outgoing modes are always excited. Then we can write $\langle\psi\rangle=\langle A^{(in)}\rangle e^{-i\omega t}
 g^{(in)}(\rho)$. By using this relation in (53) and $ F_{t\psi}=E_{0}e^{-i\omega t}$ we obtain,
 \begin{equation}
   A^{(in)}=\frac{2\pi\acute{\alpha} E_{0}}{r_{h}f(\rho)\rho^{2}g'^{(in)}(\rho)}\mid _{\rho=\rho_{b}}.
 \end{equation}
The position of particle, is given by,
\begin{equation}
  \langle\Psi(t)\rangle=\langle\psi(t,\rho_{b})\rangle= \langle A^{(in)}e^{-i\omega t}
  g^{(in)}(\rho_{b})\rangle=e^{-i\omega t}\frac{2\pi\acute{\alpha}
  E_{0}g^{(in)}(\rho_{b})}{r_{h}f(\rho)\rho^{2}g'^{(in)}(\rho)}
\mid{\rho=\rho_{b}},
\end{equation}
and then,
\begin{equation}
 \langle p(t)\rangle=e^{-i\omega t}\frac{-2im\omega\pi\acute{\alpha}
 E_{0}g^{(in)}(\rho_{b})}{r_{h}f(\rho)\rho^{2}g'^{(in)}(\rho)}\mid _{\rho=\rho_{b}}.
\end{equation}
The above relation leads us to the following result for the admittance,
\begin{equation}
\mu(\omega)=\frac{-2im\omega\pi\acute{\alpha} g^{(in)}(\rho_{b})}{r_{h}f(\rho)\rho^{2}g'^{(in)}(\rho)}\mid
_{\rho=\rho_{b}}.
\end{equation}
For conformal black hole in the zero frequency  and $\frac{r_{b}}{r_{h}}\gg1$ limits, we can get,
\begin{equation}
\mu(0)=\frac{9\acute{\alpha}m}{32\pi l^{4}T^{2}}=\frac{1}{\gamma_{0}}=t_{relax}
\end{equation}
\subsubsection{Uncharged black hole with the general mass}
Uncharged black hole, with the $f(r)$ defined through relation (7), can be investigated in the similar way to the conformal black hole in the last section. Let us begin our work with the tortoise coordinate,
 \begin{equation}
  r _{*}=\frac{l^{2}}{r_{h}}\left[\frac{\ln(\rho-1)}{(a-1)(b-1)}+\frac{a\ln(\rho-a)}{(a-b)(a-1)}
  -\frac{b\ln(\rho-b)}{(a-b)(b-1)}\right],
 \end{equation}
with $a=\frac{r_{1}}{r_{+}}$ and $b=\frac{r_{2}}{r_{+}}$.
 So, we except our solutions in the regime near the horizon has the following form,
 \begin{equation}
 \psi^{(out/in)}(t,r)\sim e^{-i\omega(t\pm r_{*}}\sim e^{-i\omega(t\pm \frac{\omega
 l^{2}\ln(\rho-1)}{r_{+}(a-1)(b-1)}}.
 \end{equation}
 In the regime (II), from the equation (54) for this kind of black hole, one can obtain,
 \begin{equation}
   g^{0}_{\upsilon}(\rho)=B_{1}+B_{2}\left[\frac{-\ln(\rho)}{ab}+\frac{\ln(\rho-1)}{(a-1)(b-1)}+
   \frac{\ln(\rho-a)}{a(a-b)(a-1)}-\frac{\ln(\rho-b)}{b(a-b)(b-1)}\right].
 \end{equation}
 By expanding (61) and comparing it with the above relation in the limit $\rho\rightarrow1$, $B_{1}$ and $B_{2}$ are
 easily derived as the following,
 \begin{equation}
  B_{1}=1\mp\frac{i\upsilon}{(a-b)}\left[\frac{\ln(1-a)}{a(a-1)}-\frac{\ln(1-b)}{b(b-1)}\right]
  \qquad,\:B_{2}=\pm i\upsilon.
 \end{equation}
 For solutions in the regime (III), we use the equation (46) and receive to the similar solutions for conformal black
 hole. As before, we expand the solutions in the regime (II) for $\rho\rightarrow\infty$ and compare them with
 solutions in the regime (III) to get coefficients $C_{1}$ and $C_{2}$,
 then we have,
 \begin{equation}
  C_{1}=B_{1}\qquad\qquad,\qquad\qquad\;C_{2}=\mp \frac{1}{\upsilon^{2}}.
 \end{equation}
 Ultimately, our asymptotic solutions, in the low frequency limit can be obtained as
 \begin{equation}
   g^{(out/in)}_{\upsilon}(\rho)\sim
   1\mp\frac{i\upsilon}{(a-b)}\left[\frac{\ln(1-a)}{a(a-1)}-\frac{\ln(1-b)}{b(b-1)}\right]
   (1+\frac{\upsilon^{2}}{2\rho^{2}})\mp\frac{i\upsilon}{3\rho^{3}}.
 \end{equation}
 Now, if we consider the case that an external force imposes on Brownian particle, we can derive the admittance by using the relation (57), therefore we get,
 \begin{equation}
   \mu(0)=\frac{2\pi\alpha^{\prime}m}{r_{+}^{2}},
 \end{equation}
 where we assumed the zero frequency and $\frac{r_{b}}{r_{+}}\gg1$ limit. In the above relation, the parameter $r_{+}$ is related to the temperature and mass of black hole through the following equation,
 \begin{equation}
  4\pi T r_{+}^{3}-3Mr_{+}^{2}=-M^{2}l^{2}.
 \end{equation}
 \subsubsection{Strings in the charged hairy black hole}
 In the case of charged hairy black hole, the method of solving the equation of motion is the same as one done in the last section. It means that we must work in three regimes that introduced  before. The first step is to derive the tortoise coordinate from the relation (28).
  The behavior of this parameter near the horizon helps to derive the solutions in the regime (I). In the next step we obtain solutions in the regime (II), through the relation (44). By expanding these solutions near the horizon and matching them with the solutions in regime (I), one can receive to the exact solutions in regime (II). The final step is to derive the asymptotic solutions from expanding the solutions in the regime (II) for $\rho\rightarrow\infty$ and comparing them with the equation (46) in regime (III). In working through the above processes for charged hairy black hole, we encountered some problems. In fact solving the integrals (28) and (44), for $f(r)$ defined through (3), is rather difficult. It seems that for this kind of black hole, we will achieve the similar relation for admittance i.e., $\mu(0)=\frac{2\pi\alpha^{\prime}m}{r_{+}^{2}}$. So, the admittance parameter will be related to the charge, mass and temperature of black hole (e.g. relation (6) for charged BTZ black hole). This statement is an opinion and we would like to confirm it at future work.
\subsection{Displacement square and the fluctuation-dissipation theorem}
This section is devoted to the computation of displacement square of the external quark and investigation of
fluctuation-dissipation theorem. The knowledge of computation of displacement square is brought in section 3 for an arbitrary background. We use this knowledge for our hairy metric backgrounds. To start, we consider the case that the fluctuating electric fields are turned off i.e., the case of free Brownian motion. According to (53)  this is,
\begin{equation}
  \frac{r_{h}f(\rho)\rho^{2}\partial_{\rho}\psi}{2\pi \alpha^{\prime}}=0\qquad\qquad\mathrm{at}\qquad\rho=\rho_{b}.
\end{equation}
We know that $\psi(t,r)$ can be written as the sum of outgoing and ingoing modes, (recall eq. (33)),
\begin{equation}
\psi(t,\rho) =\eta\left[g^{(out)}(\rho)+\delta\:g^{(in)}(\rho)\right]e^{-i\omega t}\,.
\end{equation}
From (67), it is governed that,
\begin{equation}
\delta=-\frac{g^{(out)'}}{g^{(in)'}}.
\end{equation}
$\delta$ is a pure phase and $g^{(out)}=g^{(in)*}$. For hairy black holes with the metric background (2), we have
$\eta=\sqrt{\frac{4\pi\alpha^{\prime}}{\omega r_{+}^{2}}}$ and for uncharged black hole, to leading order in frequency one can find that $\delta=1$ and then $\left|g^{(out)}+\delta\:g^{(in)}\right|^{2}=1+O(\upsilon^{2})$. So we can deduce the late-time behavior of the displacement square from the low frequency limit of (37) i.e.,
\begin{equation}
 S^{2}(t)=\frac{16\alpha^{\prime}}{\beta r_{+}^{2}}\int_{0}^{\infty}d\omega\frac{\sin^{2}(\frac{\omega t}{2})}{\omega^{2}}\sim
 \frac{4\pi\alpha^{\prime}}{\beta r_{+}^{2}}t\,.
\end{equation}
Thus, we get that the diffusion constant defined as in (17) is given by,
\begin{equation}
  D=\frac{2\pi\alpha^{\prime}T}{r_{+}^{2}},
\end{equation}
In the special case of conformal black hole, the relation (70) reduces to,
\begin{equation}
 S^{2}(t)\sim \frac{9\alpha^{\prime}}{16l^{4}\pi T}t,
\end{equation}
then $D=\frac{9\alpha^{\prime}}{32l^{4}\pi T}$. For general mass in uncharged black hole, the horizon radius depends on mass and temperature of black hole, so in this case, the diffusion constant has a dependance on mass and temperature of black hole.\\
At the end of this section, we turn our attention to investigate the fluctuation-dissipation theorem (16). In
order to check this, we compute the random force autocorrelation in (23) to obtain the coefficient $\kappa_{0}$.
 From (36) we can derive the correlator of momentum as,
\begin{equation}
\langle :p(t)p(0):\rangle=\int_{0}^{\infty}\frac{d\omega}{2\pi}\frac{2m^{2}\omega^{2}|\eta^{2}|\cos(\omega
t)}{e^{\beta\omega}-1}=\int_{-\infty}^{\infty}\frac{d\omega}{2\pi}\frac{4\pi Tm^{2}\alpha^{\prime}}
{r_{+}^{2}}\frac{\beta|\omega|e^{-i\omega
t}}{e^{\beta\omega}-1}.
\end{equation}
Therefore,
\begin{equation}
  I_{p}(\omega)=\frac{4\pi Tm^{2}\alpha^{\prime}}
{r_{+}^{2}}\frac{\beta|\omega|}{e^{\beta\omega}-1}.
\end{equation}
By combining this with (65), at leading order one find that,
\begin{equation}
 I_{R}(\omega)=\frac{r_{+}^{2}T}{\pi\alpha^{\prime}}+O(\omega),
\end{equation}
 this gives the coefficient $\kappa_{0}=\frac{r_{+}^{2}T}{\pi\alpha^{\prime}}$, which in the special case of conformal black hole is equal to $\kappa_{0}=\frac{64\pi l^{4}T^{3}}{9\alpha^{\prime}}$. By checking the relation (16), we can see that the fluctuation-dissipation theorem holds in the plasma where the corresponding gravity is three dimensional uncharged hairy black hole. In the case of charged hairy black hole likewise the uncharged black hole, we think that this theorem is also hold. Because it seems that in the low frequency limit the $g(\rho)$  function for charged black hole has a similar behavior with $g(\rho)$  function for uncharged black hole.
 \section{Conclusion}
In this paper, we obtained the normalized asymptotic solutions (including outgoing and ingoing modes) to the equation of motion of uncharged hairy black hole at low frequencies. By using those solutions, we derived the response function and correlation function for uncharged black hole in general mass and special mass $M=\frac{3B^{2}}{l^{2}}$ separately. We found that the admittance and diffusion constant are dependent on the scalar parameter $B$ and mass of black hole trough the horizon radius. We proved that the fluctuation-dissipation theorem holds in the plasma where the corresponding gravity is three dimensional uncharged hairy black hole. In the case of charged hairy black hole, we could not get an explicit solution to the equation of motion, but we think that its behavior of asymptotic solution is similar to the uncharged case at the low frequency limit. It means that the dependence of admittance and diffusion constant on horizon radius is as before and in the case of charged black hole related to the scalar parameter, charge and mass of black hole. This statement is only a comment and should be confirmed in the future works.\\\\

\end{document}